\documentclass{optica-article}

\journal{opticajournal} 

\articletype{Research Article}

\usepackage{lineno}
\usepackage{amsmath}
\usepackage{mathdots}
\usepackage{braket}
\usepackage{subfigure}
\usepackage{graphicx}
\usepackage{dcolumn}
\usepackage{bm}

\begin{document}

\title{A continuous cold rubidium atomic beam with enhanced flux and tunable velocity}

\author{Shengzhe Wang,\authormark{1,2} Zhixin Meng,\authormark{1,2,3},Peiqiang Yan,\authormark{1,2} Yuanxing Liu,\authormark{3} and Yanying Feng \authormark{1,2,*}}

\address{\authormark{1}State Key Laboratory of Precision Measurement Technology and Instruments, Tsinghua University, Beijing, China.\\
\authormark{2}Department of Precision Instrument, Tsinghua University, Beijing, China.\\
\authormark{3}Beijing Institute of Aerospace Control Devices, Beijing, China.}

\email{\authormark{*}yyfeng@tsinghua.edu.cn} 


\begin{abstract*} 
We present a cold atomic beam source based on a two-dimensional (2D)+ magneto-optical trap (MOT), capable of generating  a continuous cold beam of ${}^{87}$Rb atoms with a flux up to $4.3\times10^9 \, \rm{s^{-1}}$, a mean velocity of $10.96(2.20)\,\rm{m/s}$, and a transverse temperature of $16.90(1.56)\,\rm{\mu K}$. Investigating the influence of high cooling laser intensity, we observe a significant population loss of atoms to hyperfine-level dark states. To account for this, we employ a multiple hyperfine level model to calculate the cooling efficiency associated with the population in dark states, subsequently modifying the scattering force. Simulations of beam flux at different cooling and repumping laser intensities using the modified scattering force are in agreement with experimental results. Optimizing repumping and cooling intensities enhances the flux by 50\%. The influence of phase modulation on both the pushing and cooling lasers is experimentally studied, revealing that the mean velocity of cold atoms can be tuned from $9.5\,\rm{m/s}$ to $14.6\,\rm{m/s}$ with a phase-modulated pushing laser. The versatility of this continuous beam source, featuring high flux, controlled velocity, and narrow transverse temperature, renders it valuable for applications in atom interferometers and clocks, ultimately enhancing bandwidth, sensitivity, and signal contrast in these devices.

\end{abstract*}

\section{Introduction}

Continuous cold atomic beams are significant in various applications, including quantum sensing \cite{Xue2015,Meng2020a,Kwolek2022}, atomic frequency standard \cite{Elgin2019,Meng2020}, and fundamental research, such as Bose-Einstein condensation (BEC) and the investigation of precise atomic spectra \cite{Domenico2001}. In comparison to pulsed cold atomic clouds, continuous cold atomic beams offer advantages in terms of increased data rates and the elimination of dead time, mitigating aliasing noises induced by the Dick effect  \cite{Joyet2012, Xue2015,Meng2020a,Kwolek2022}. Additionally, when compared with thermal atomic beams\cite{Gustavson2000a,Durfee2006b}, cold atomic beams provide a slower mean velocity and narrower velocity distribution of atoms. These characteristics contribute to extended interrogation times and enhanced fringe contrasts in interferometers and clocks.

Diverse continuous cold atomic beams have been successfully demonstrated and can be classified into two primary methods. The first involves decelerating a thermal atomic beam with radiation pressure by using  Zeeman slowers \cite{Phillips1982,Paris-Mandoki2014,YangWei2014}, frequency-chirped lasers \cite{Paris-Mandoki2014,Ertmer1985,Truppe2017}, isotropic light slowing \cite{Ketterle1992}, or white-light cooling \cite{Park2002,Zhu1991}. Cold atomic beams generated through these approaches typically exhibit high atom fluxes. However, they often result in an elevation of the transverse temperature of atoms, consequently leading to a reduction in beam brightness and phase space density.

The alternative approach involves cooling atoms within a vapor chamber using magneto-optical traps (MOTs) and subsequently ejecting the atoms into a collimated beam. Various configurations of magnetic and optical fields have been demonstrated, employing two-dimensional (2D) MOTs \cite{Riis1990,Swanson1996,Weyers1997,Berthoud1998,Schoser2002,Ramirez-Serrano2006,Castagna2006,Muller2007,Kellogg2012,Rathod2013,Chanu2016}, 2D+MOTs \cite{Dieckmann1998,Wang2003,Conroy2003,Ovchinnikov2005,Catani2006,Xie2022}, three-dimensional (3D) MOTs or low-velocity intensity sources (LVIS) \cite{Lu1996,Arlt1998,Camposeo2001,Park2002,Kohel2003,Wang2011a}. These methods offer a trade-off between atom flux and temperature compression while typically resulting in a low longitudinal velocity of cold atoms and a reduced background of thermal atoms. Additionally, multi-stage cooling involving MOTs and optical molasses has been employed to generate cold atomic beams with exceptional performance \cite{Kwolek2020,Devenoges2012,Roy2012,Berthoud1999}.

In this study, we present and characterize a cold atomic beam source utilizing a 2D+MOT. The source demonstrates a continuous high atomic flux, reaching up to $4.3\times10^9 \, \rm{s^{-1}}$, with a mean atom velocity of $10.96(2.20)\,\rm{m/s}$, a transverse temperature of $16.90(1.56)\,\rm{\mu K}$ and a longitudinal velocity distribution of $2.92(1.24)\,\rm{m/s}$ (full width at half maximum, FWHM). The influence of the repumping laser on beam flux is investigated theoretically and experimentally. Previous models for cooling alkali-metal atoms in 2D+MOTs were based on the two-level approximation, neglecting the impact of the repumping laser \cite{Ovchinnikov2005,Wang2011a,Metcalf1999,Huntington2023}. However, the influence of hyperfine-level dark states becomes notable, particularly when the cooling laser intensity is high\cite{Xie2022,Lu1996,Wang2011a}. We develop a modified 2D+MOT model and conduct an analysis of atomic dynamics spanning multiple hyperfine levels to account for the cooling efficiency influenced by atoms in dark states, determining the repumping intensity requirements of the 2D+MOT at varying cooling intensity levels. Additionally, we investigate the effects of white-color cooling and pushing lasers in the 2D+MOT configuration to fine tune the performance of the atomic beam, and realize an adjustment of the mean velocity from $9.5\,\rm{m/s}$ to $14.6\,\rm{m/s}$.

\section{Experimental setup}

\begin{figure}[b]
\centering
\includegraphics[width=0.65\linewidth]{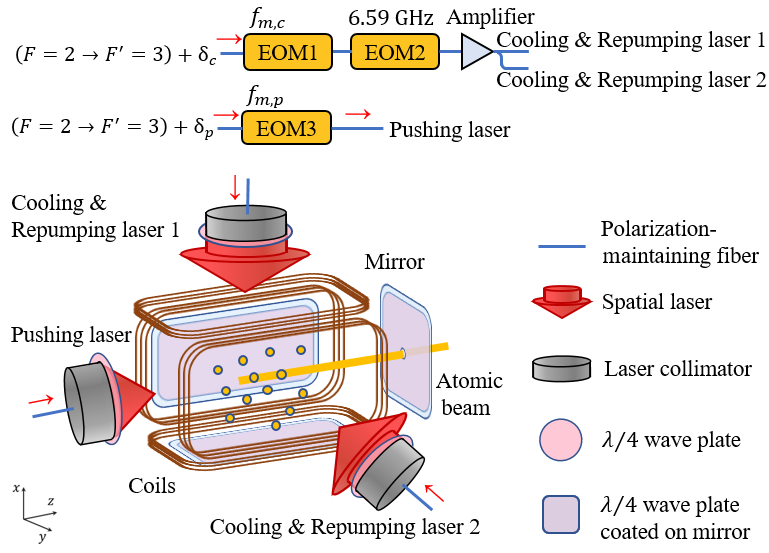}
\caption{\label{fig1} The schematic diagram of the cold atomic beam source. The carrier lasers are generated from a frequency shift laser system with the detuning $\delta_c\approx-4\Gamma$ and $\delta_p\approx-4\Gamma$ from $F=2 \to F^{'} =3$, which are optimized separately. EOM1 and EOM3 are used to modulate the cooling laser and the pushing laser at $f_{m,c}$ and $f_{m,p}$, thereby generating white lasers with broader spectra. EOM2 is used to generate the a repumping frequency on the cooling laser. Lasers are then sent to the collimator through polarization-maintaining fibers.}
\end{figure}

A schematic of the cold atomic beam source of $^{87}$Rb is illustrated in Fig.~\ref{fig1}. The setup comprises two pairs of counterpropagating cooling laser beams passing through an atomic vapor within a vacuum chamber. These laser beams propagate perpendicularly to each other along the x and y axes. Passing through collimators and $\lambda⁄4$ wave plates, the beams transform into circular-polarized beams. Mirrors, coated with $\lambda⁄4$ wave plate films, reflect the beams to provide $\sigma^+/\sigma^-$ circular polarization. A gradient magnetic field of approximately 10 Gauss/cm along the x and y axes is generated by two pairs of anti-Helmholtz coils.

A pushing laser beam, following a similar path, passes through a collimator and a $\lambda⁄4$ wave plate. This beam is directed through the atomic vapor along the z-axis and is partially reflected back by a mirror with a centrally drilled 1-mm diameter pinhole. Consequently, the atoms along the z-axis undergo cooling, and simultaneously, the cold atoms are propelled out of the pinhole. The mirror for the pushing laser is also coated with a $\lambda⁄4$ wave plate film.

Both cooling and pushing lasers operate at typical  frequencies with a $4\Gamma$ red detuning (where $\Gamma = 6.06\,\rm{MHz}$ is the natural linewidth) from the $F=2 \to F^{'} =3$ transition. The specific detuning has been separately optimized during experiments. These lasers are phase-modulated by electro-optic modulators (EOMs) to generate white lasers with broader spectra, as shown in Fig.~\ref{fig1}. EOM1 and EOM3 phase-modulate the cooling laser and the pushing laser, respectively, while EOM2, driven by a 6.59 GHz RF source,  phase-modulate the cooling laser to generate the repumping laser. The intensity ratio between the cooling and repumping lasers can be fine-tuned by adjusting the modulation depth of EOM2.

\begin{figure}[h]
\centering
\includegraphics[width=0.6\linewidth]{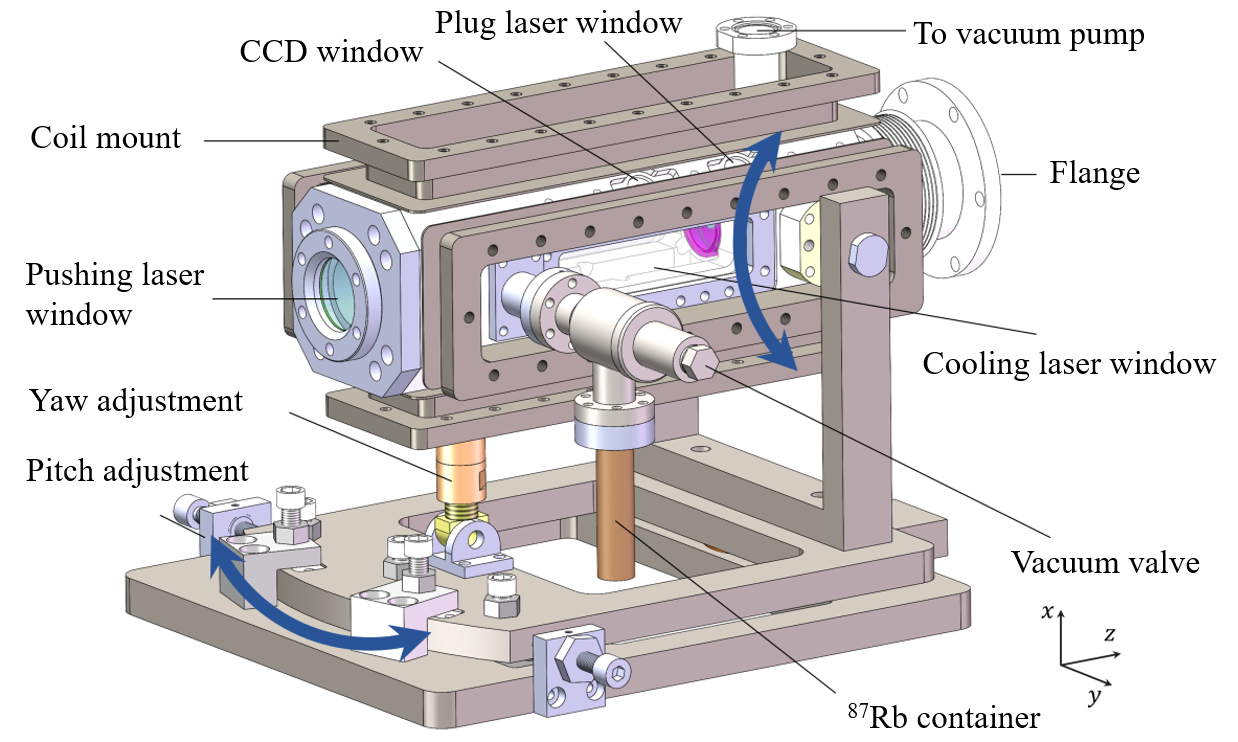}
\caption{\label{fig2} The diagram of the source’s vacuum chamber and the peripherals. The deep blue arrows depicted the adjustment direction of the MOT.}
\end{figure}

The direction of the atomic beam propagation can be precisely adjusted by rotating the 2D+ MOT setup around the x-axis and y-axis, as shown in Fig.~\ref{fig2}. Fine adjustments to the yaw angle of the setup are achieved using two adjustment threads, while a dedicated thread allows for precise modifications to the pitch angle. The rotation is facilitated by a flanged bellow connection between the MOT and the scientific chamber. This direction adjustment structure proves highly beneficial, particularly in applications involving dual counterpropagating atomic beams, where collinearity between the beams is a requirement. The thermal atom vapor is generated by heating the Rb container. A  window is designed for monitoring the formation of atomic beams. To mitigate the impact of the pushing laser on the atomic beam, the atoms propagate with a parabolic trajectory.  A state-preparation laser beam pumps the atoms to $\ket{F=1,m_F=0}$ state of $^{87}$Rb, after the atoms exit the 2D+MOT.

\section{Results and discussion}

\subsection{Velocity distributions and Flux}

The cold atomic beam is detected through fluorescence from spontaneous emission. The atoms pass through counter-propagating detection lasers resonant with the $F=2 \to F^{'} =3$ transition, continually scattering photons in random direction. Photons in a specific direction are then focused by an optical system onto a photo-multiplier tube (PMT, H7422-50, Hamamatsu, Japan).

\begin{figure}[t]
\centering
\includegraphics[width=0.6\linewidth]{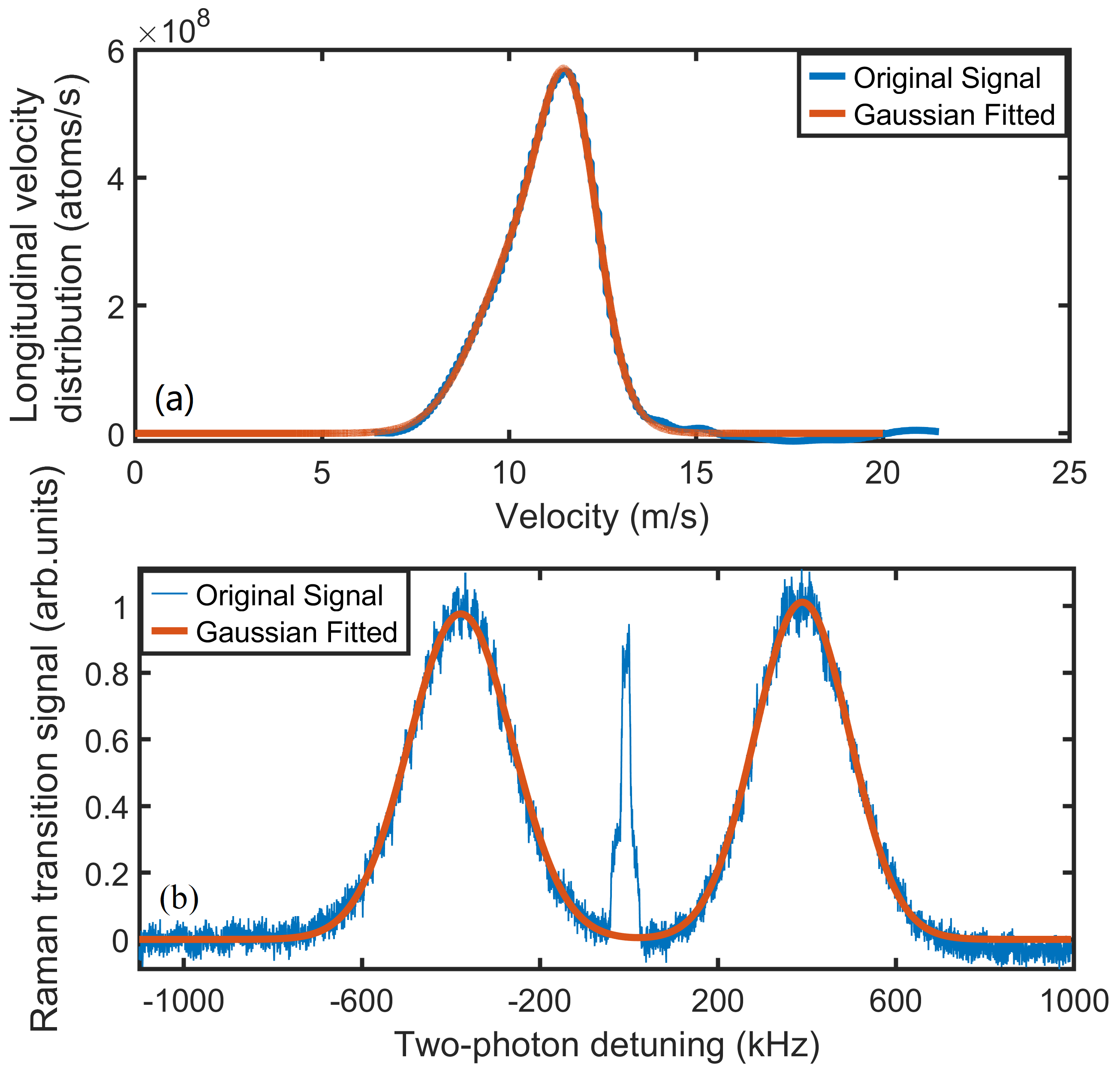}
\caption{\label{fig3} Longitudinal velocity distribution from a TOF signal (a), Doppler-sensitive stimulated Raman transitions (b) based on the cold atomic beam source, with intensities corresponding to a $\pi$ Raman pulse. A fit of the Doppler-sensitive distribution to a Gaussian reveals a FWHM of $245\,\rm{kHz}$, corresponding to the transverse temperature of $16.90(1.56)\,\rm{\mu K}$. The Raman laser contains two counter-propagating laser beams. Lasers of both directions have two frequency components, leading to the Doppler-free transition in the middle range in (b).}
\end{figure}

The velocity and flux of the continuous atomic beam are determined using the  time-of-flight (TOF) method. This method involves introducing a plug laser, resonant with $F=2 \to F^{'} =3$, at the exit of the MOT while simultaneously recording the falling edge of the fluorescence signal from atoms in the $F=2$ state. The atoms travel a distance of 64.1 cm before reaching the detection laser. Fig~\ref{fig3}(a) shows a typical longitudinal velocity spectrum derived from the TOF signal.   

Doppler-sensitive stimulated Raman transitions, depicted in Fig.~\ref{fig3}(b),  are employed to determine the transverse velocity distribution of the atomic beam \cite{Xue2015,Kwolek2020}. The lower limit for transverse temperature measurement is $18\,\rm{nK}$, determined by  the Fourier transform of a 1-mm width profile of the Raman beam. The Raman laser is red-detuned by $1\,\rm{GHz}$ from the ${}^{87}$Rb $D_2$ line transition to avoid single-photon resonant excitation. Positioned approximately 30 mm from the MOT exit, the Raman beam is tilted slightly away from the perpendicular direction relative to the atomic beam, in order to distinguish the Doppler-sensitive spectrum from the residual Doppler-insensitive spectrum.

The cold atomic beam is measured to exhibit a maximum flux of  $4.3\times10^9 \, \rm{s^{-1}}$ after optimizing the intensity, polarization, frequency, propagation direction of cooling laser, along with the magnitude field gradient. The  longitudinal velocity distribution is determined  to be $2.92(1.24)\,\rm{m/s}$ (FWHM), with the most probable velocity of $10.96(2.20)\,\rm{m/s}$. The transverse temperature is calculated as $16.90(1.56)\,\rm{\mu K}$ based on the linewidth of the Doppler-sensitive Raman spectrum. The anomalously low temperature, below the Doppler cooling limit, may be attributed to polarization gradient cooling (PGC) \cite{Metcalf1999}. Considering the imperfect purity of the cooling lasers polarization, both $\sigma^+/\sigma^-$ and $lin\perp lin$ configuration could contribute to PGC. The transverse size expansion of the atom beams and the numerical aperture inhomogeneity at different field view of the fluorescence collection optical system could underestimate the transverse temperature.

\subsection{Flux’s dependence on repumping intensity at high cooling intensity}

A 2D+ MOT beam is typically modeled with a two-energy-level structure \cite{Metcalf1999,Wang2003,Ovchinnikov2005,Xie2022}. However, some atoms escape the transition cooling cycles and become uncoolable due to the presence of multiple hyperfine structure energy levels. This phenomenon becomes more pronounced at higher cooling laser intensities, resulting in a decrease in the flux of the 2D+MOT beam. In this section, we discuss the repumping intensity needed for the 2D+MOT at different cooling levels, considering the impact of multiple hyperfine structure levels.

\begin{figure}[h]
\centering
\includegraphics[width=0.45\linewidth]{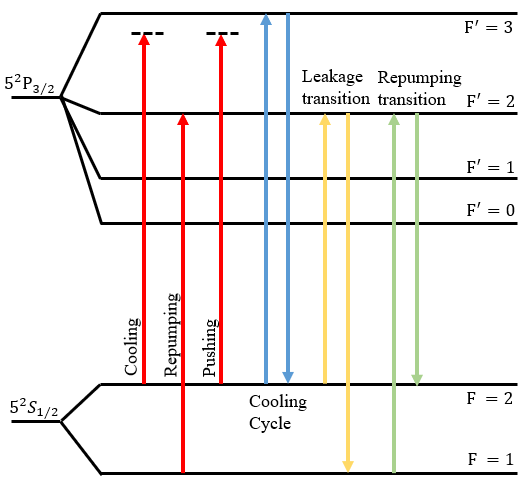}
\caption{\label{fig4} The energy levels of the ${}^{87}$Rb $D_2$ line, with the frequencies of cooling, repumping and pushing lasers (red). The atoms’ transitions during the cooling process includes the cooling transition cycle (blue), the leakage transition (yellow), and the repumping transition (green).}
\end{figure}

The six-level generalization of the optical Bloch equations to the ${}^{87}$Rb $D_2$ line is difficult to treat both analytically and numerically. We simplify the transition rate $R_{FF^{'}}$ among the specific hyperfine levels as the product of the degenerate scattering rate $R_{sc}$ and the relative transition strength factors $S_{FF^{'}}$ \cite{Steck2003Rubidium8D},

\begin{equation}
R_{FF^{'}} = S_{FF^{'}}R_{sc} = S_{FF^{'}}\frac{\Gamma}{2}\frac{s}{1 + s + \left( {{2\delta_{FF^{'}}}/\Gamma} \right)^{2}}
\end{equation}
where $\Gamma$ is the natural line width of the ${}^{87}$Rb $D_2$ line, $s=I_0⁄I_{sat}$  is the saturation intensity factor, $I_0$ is the laser intensity, $I_{sat}$ is the saturation intensity, $\delta_{FF^{'}}$ is the laser detuning from the transition $F\to F^{'}$. The transition rates from $F=2$ are $R_{c,~23} = 282.80 \times 10^{3}~\rm{{s}^{- 1}}$, $R_{c,~22} = 1.18 \times 10^{3}~\rm{{s}^{- 1}}$, $R_{c,~21} = 0.087 \times 10^{3}~\rm{{s}^{- 1}}$, driven by a cooling laser with $-4\Gamma$ detuning from the $F=2 \to F^{'} =3$ transition and $s=10$ intensity factor. Because $R_{c,21} \ll R_{c,22}$ , we can ignore the atoms in $F^{'}=1$ and $F^{'}=0$. The dynamics of the atoms system can be described by a set of differential equations,

\begin{equation}\left\{
\begin{aligned}
\frac{dn_{3^{'}}}{dt} &= R_{c,23^{'}}n_{2} - \Gamma n_{3^{'}} \\
\frac{dn_{2^{'}}}{dt} &= R_{c,22^{'}}n_{2} + R_{re,12^{'}}n_{1} - \Gamma n_{2^{'}} \\
\frac{dn_{2}}{dt} &= \Gamma n_{3^{'}} + \frac{1}{2}\Gamma n_{2^{'}} \\
\frac{dn_{1}}{dt} &= \frac{1}{2}\Gamma n_{2^{'}}
\end{aligned}\right.
\end{equation}
where $n_{F}$ and $n_{F^{'}}$ are the populations in $\ket{F}$ or $\ket{F^{'}}$ .  $R_{c,FF{'}}$ and $R_{re,FF{'}}$ are the transition rate driven by the cooling laser and the repumping laser. We transform the differential equations into the equations that describe the population in the cooling transition cycle $n_{cycle}$, and out of the cooling transition cycle $n_{out}$.

\begin{equation}
\begin{aligned}
\frac{dn_{cycle}}{dt} &= \frac{dn_{3^{'}}}{dt} + \frac{dn_{2}}{dt} = R_{c,23^{'}}n_{2} + \frac{1}{2}\Gamma n_{2^{'}} \\
\frac{dn_{out}}{dt}   &= \frac{dn_{2^{'}}}{dt} + \frac{dn_{1}}{dt} = R_{c,22^{'}}n_{2} + R_{re,12^{'}}n_{1} - \frac{1}{2}\Gamma n_{2^{'}}
\end{aligned}
\end{equation}

When the system is in a steady-state with both cooling and repumping laser, there is a balance between the rate of atoms escaping from and being repumped back to the cooling cycle. Ignoring the population in the excited level, there is 

\begin{equation}
\frac{dn_{cycle}}{dt}=\frac{dn_{out}}{dt}
\end{equation}

\begin{equation}
\eta = \frac{n_{cycle}}{n_{cycle}+n_{out}} = \frac{R_{re,12^{'}}}{R_{re,12^{'}}+R_{c,23^{'}}-R_{c,22^{'}}}
\end{equation}
where $\eta$ represents the ratio of atom population in the cycle and functions as an indicator of cooling efficiency. This is because only atoms within the cooling cycle resonate with the cooling laser. When all atoms are in the cooling cycle, $\eta=1$.
The scattering force from the trapping laser beams on the atoms is given by \cite{Wang2011}

\begin{equation}
F_c = \hbar k \frac{\Gamma}{2} \frac{s_{c}}{1+s_{c}+(2\delta_{c,23^{'}} /\Gamma)^2}
\end{equation}
where $\hbar k$ is the photon momentum, $s_c$ is the saturation factor of the cooling laser, $\delta_c$ is the detuning between the cooling laser and the atoms’ transition. Taking the cooling efficiency in Eq.(5) into consideration, the scattering force can be expressed as

\begin{equation}
\begin{aligned}
F_{c,\eta} &= \eta \times F_c = \frac{R_{re,12^{'}}}{R_{re,12^{'}}+R_{c,23^{'}}-R_{c,22^{'}}}  \times \hbar k \frac{\Gamma}{2} \frac{s_{c}}{1+s_{c}+(2\delta_{c,23^{'}} /\Gamma)^2}
\end{aligned}
\end{equation}

$F_{c,\eta}$ is a relatively accurate expression of the scattering force by the cooling laser on the atoms with the consideration of atoms in dark states. Referring to Eq.(7), an increase in cooling laser power ($s_c$) leads to a rise in the cooling rate ($F_c$). However, the cooling efficiency ($\eta$) drops, resulting in a decrease in the atomic beam's flux. This effect is particularly noticeable when the repumping laser intensity is below $I_{sat}$.

We formulated a numerical simulation model using the Monte Carlo method to assess the flux of the 2D+MOT cold beam source. Initial atoms are created with random three-dimensional positions and velocities, following a normal distribution. These atoms are treated as classical particles and are subjected to acceleration from the six scattering forces outlined in Eq.(7). The flux is determined by counting the atoms passing through the 1-mm hole per second.

An experiment was conducted employing  identical parameters for cooling and repumping lasers, with variations introduced in the total laser power and the ratio between the cooling and repumping lasers. The total laser power was adjusted using the optical amplifier, while the ratio was manipulated through EOM2, as depicted in the Fig.~\ref{fig1}. Calibration of these parameters was accomplished using a Fabry–Perot cavity and an optical power meter. Notably, EOM1 and EOM3 were deactivated  during this experiment. 

\begin{figure*}[h]
\centering
\includegraphics[width=1\linewidth]{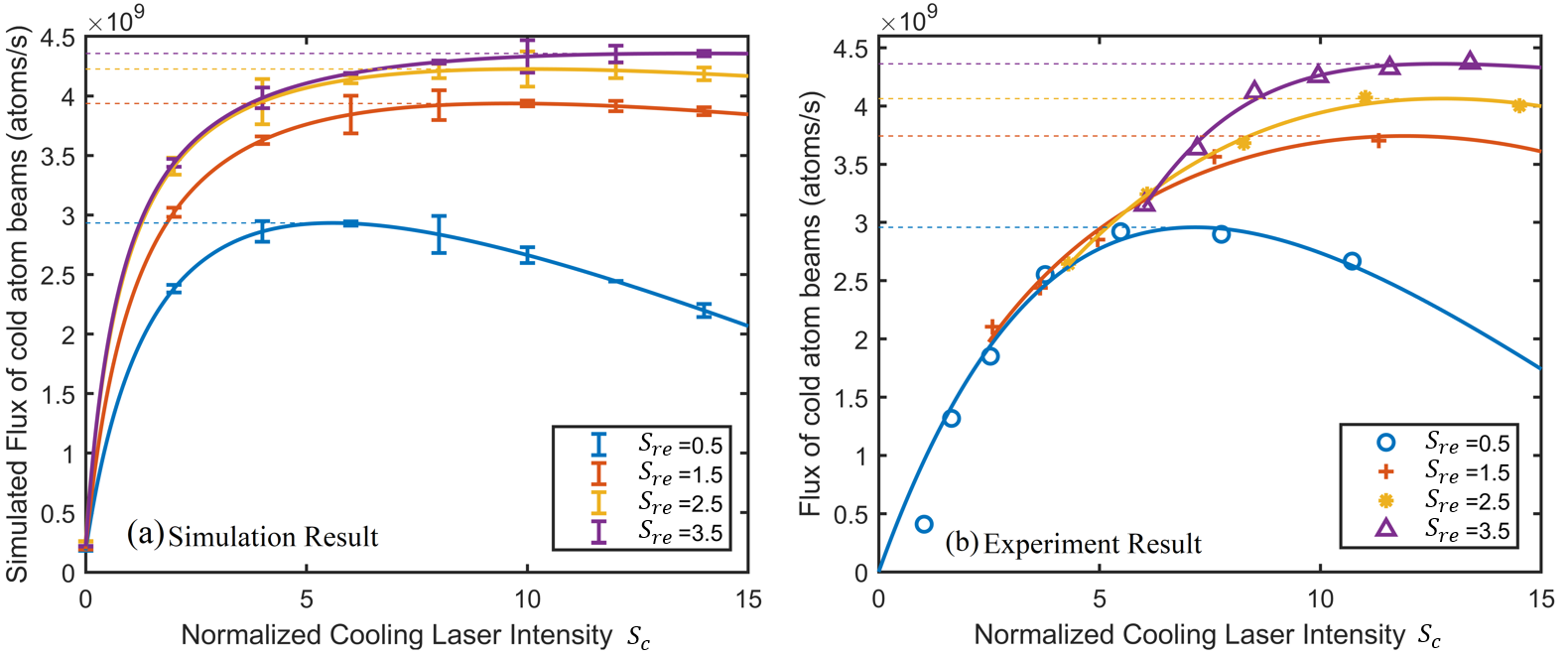}
\caption{\label{fig5} Flux’s dependence on cooling laser intensity at different repumping intensity. Results are from the simulation(a) and experiments (b). The intensities are normalized with $S_c=I_c⁄I_0$ , $S_{re}=I_{re}⁄I_0$ , where $I_c$ and $I_{re}$ are intensities of cooling and repumping laser, $I_0=1.67 \rm{mW⁄cm^2}$   is the saturation intensity for ${}^{87}$Rb atoms. The solid lines are the fits of the data points to Eq.(7), and the dash lines indicate the maximum flux at particular repumping laser intensity. Error bars in the simulation results are calculated from 10-times Monta-Carlo simulation.}
\end{figure*}

Fig.~\ref{fig5}(a) and (b) present the simulation and experimental results, respectively. Both indicate that the atomic beam's flux increases with escalating cooling laser intensity and reaches saturation, when the repumping laser intensity is high. However, under low repumping laser intensity conditions, particularly below $I_{sat}$ (e.g. $S_{re}=0.5$), the flux hits a  certain threshold as the cooling laser intensity rises and subsequently declines. The experimental results are consistent with the simulation results derived from the cooling efficiency-modified 2D+MOT model, as demonstrated by specific features in Fig.~\ref{fig5}. 

In a typical MOT setup with a low-intensity repumping laser, the atom flux increases more slowly or may even decrease as the cooling laser intensity rises, as shown by the blue lines in Fig.~\ref{fig5}. This happens because the growing cooling laser intensity reduces cooling efficiency. The higher cooling intensity causes more atoms to escape from the $F=2 \to F^{'}=3$ cooling transition cycle, lowering the number of atoms resonating with the cooling laser. Hence, there might be an optimal cooling laser intensity, especially when the repumping laser intensity is limited.

Increasing the repumping laser intensity in the MOT results in a higher flux when optimizing cooling intensities. In our study, enhancing the repumping laser intensity factor from 0.5 to 3.5 improves the beam flux by approximately 50\%. Additionally, with an increase in repumping intensity, the optimal cooling intensity also rises, suggesting that higher repumping intensity facilitates more effective interaction with the atoms. This is attributed to the compensatory effect of the high-intensity repumping laser, countering the decrease in cooling efficiency induced by the high-intensity cooling laser.

\subsection{White-color laser cooling experiments}

Laser cooling utilizing white-color lasers presents a viable method for enhancing beam flux and adjusting the velocity of a cold atomic beam. This approach holds promise for improving fringe contrast and the flexibility of an atom interferometer. White-color lasers, characterized by their widened frequency spectrum, have been investigated for their potential impact on MOTs, serving various purposes as reported in previous studies \cite{Li2015, Park2012, Barbiero2020, Lee:17}. In this section, we investigate the effects of white-color cooling and pushing lasers on a 2D+MOT cold atomic beam. The broadened frequency spectrum of the pushing laser and the cooling laser are achieved through phase modulation of the lasers using EOM1 and EOM3, respectively, as illustrated in Fig.~\ref{fig1}.

\begin{figure}[h]
\centering
\includegraphics[width=0.95\linewidth]{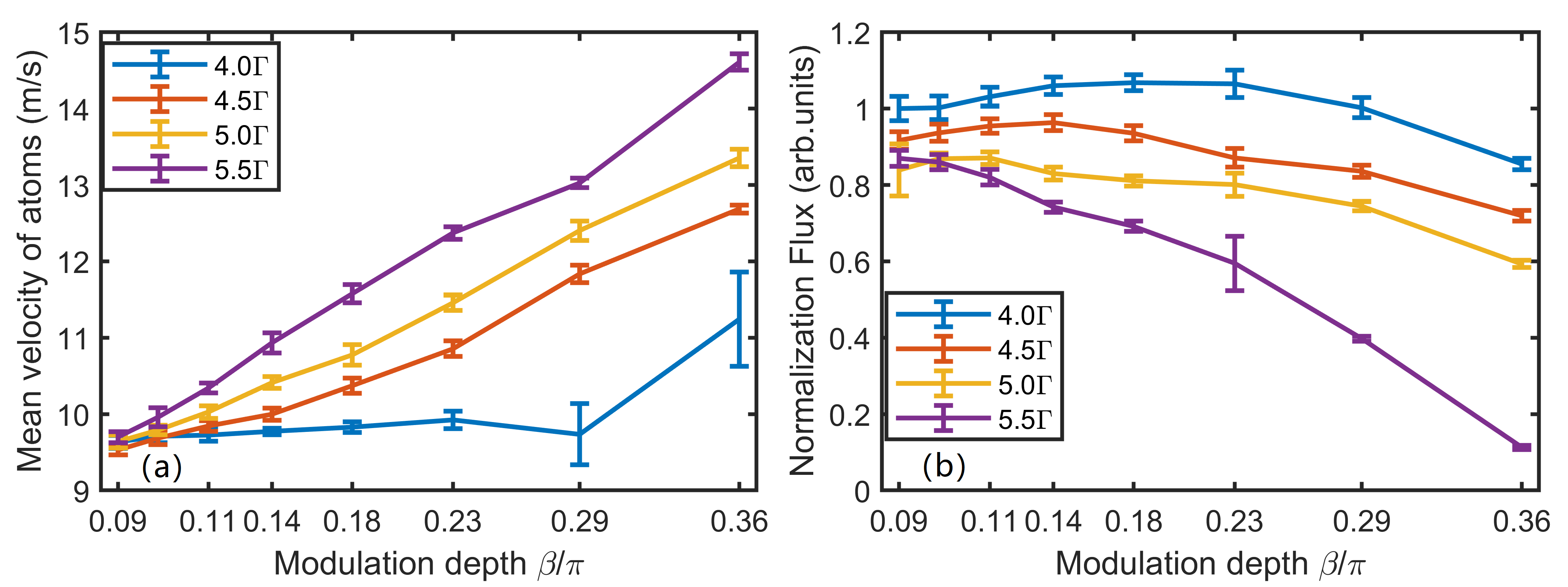}
\caption{\label{fig6} Mean velocity (a) and Flux (b) dependence of the cold beam source on the phase modulation of the pushing laser. The modulation depth is normalized by $\pi$. The pushing laser have a frequency with about $4\Gamma$ red detuning. The sidebands at different frequencies of $4.0\Gamma$(blue), $4.5\Gamma$(orange), $5.0\Gamma$(yellow), $5.0\Gamma$(purple) was generated by EOM3 in Fig.~\ref{fig1}. Error bars represent the deviations calculated from three separate experiments.}
\end{figure}

In Fig.~\ref{fig6}(a) and (b), the dependencies of the mean velocity and flux of a 2D+MOT atomic beam on the phase modulation depth of EOM3 are shown for different modulation frequencies. A RF signal is employed to modulated the pushing laser, while the cooling laser is not modulated during this experiment. The modulation depth is tuned by the power of the RF signal, which can change the optical intensity distribution among the carrier and sidebands of the pushing laser. The optical intensity of the pushing laser is $2.55\,\rm{mW/cm^2}$ ($1.52I_0$), while that of the cooling laser is $10.02\,\rm{mW/cm^2}$ ($6I_0$). The modulation depth is normalized by $\pi$, and the beam flux is normalized by the flux without white-color modulation of the pushing laser. For a specific modulation frequency, exemplified with a red detuning of $4.0\Gamma$ illustrated with blue lines, the mean velocity of the beam increases while the beam flux decreases with the modulation depth. 

It is observed that the mean velocity increases more rapidly as the modulation frequency rises. This phenomenon may be attributed to the fact that, after phase modulation, the pushing laser exhibits a blue detuning, consequently elevating the mean velocity. Additionally, the intensity of the pushing laser at a $4\Gamma$ red detuning decreases as the modulation depth increases, resulting in a reduction of the cooling effect and a subsequent decline in flux, as depicted in Fig.6(b). The achievable range of the mean velocity extends from $9.7\,\rm{m/s}$ to $14.6\,\rm{m/s}$  for a pushing laser red detuned to $5.5\Gamma$ when the modulation depth varies from 0.09 to 0.36. However, in this case, the normalized beam flux decreases to 11\% of the original flux without white-color modulation of the pushing laser. When the pushing laser is red-detuned to $5.0\Gamma$, as indicated by the orange lines in Fig.~\ref{fig6}, the tuning range of the beam mean velocity linearly spans from $9.6\,\rm{m/s}$ to $13.4\,\rm{m/s}$, and the beam flux decreases to 60\% of the original flux. This configuration proves more practical for applications requiring precise velocity tuning of an atomic beam.

\begin{figure}[t]
\centering
\includegraphics[width=0.7\linewidth]{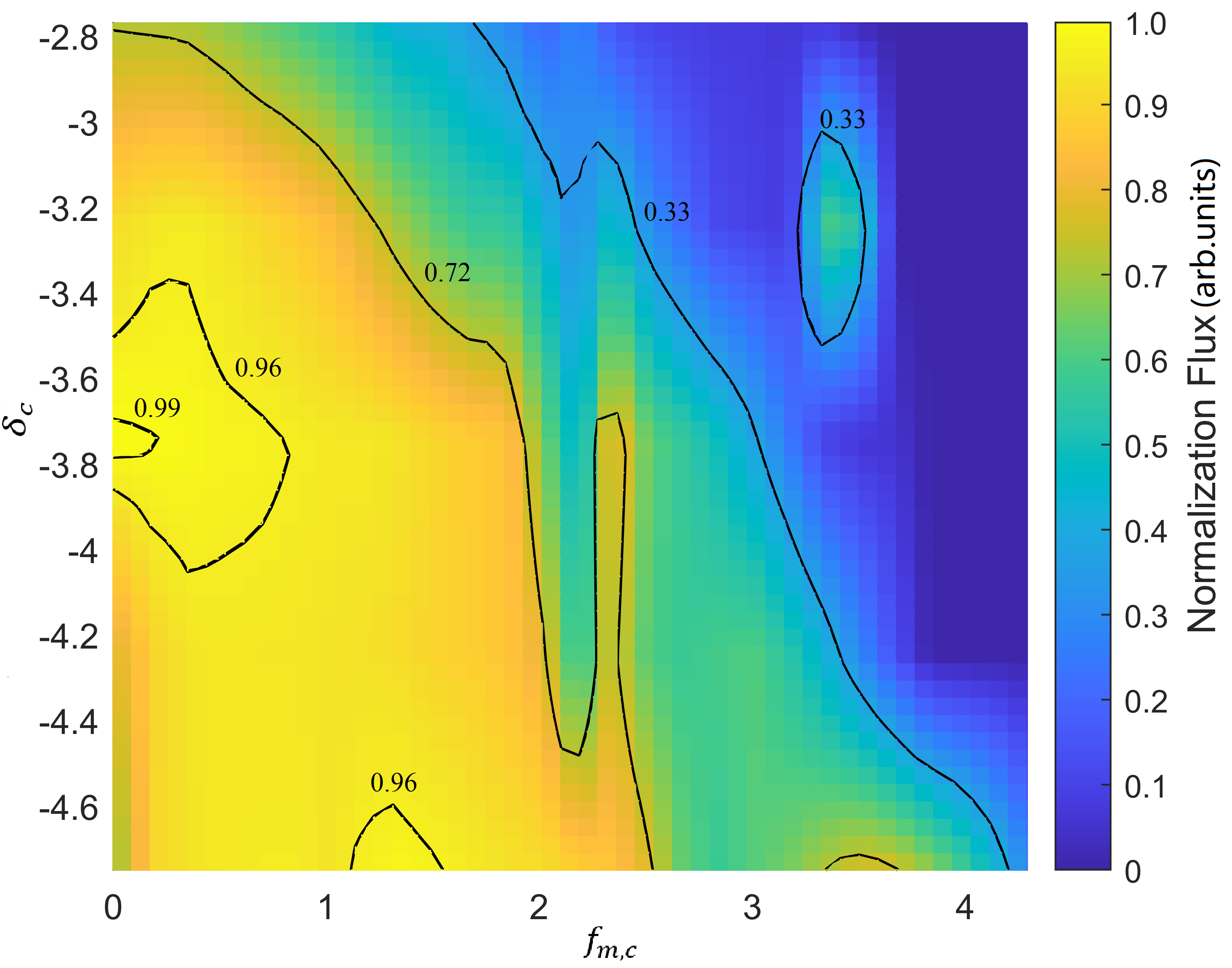}
\caption{\label{fig7} Flux as a function of carrier and modulation frequencies. $\delta_c$ is the detuning of cooling laser without phase-modulation, relative to $F=2 \to F^{'}=3$ transition. $f_{m,c}$ is the modulation frequency of the cooling laser. The detuning and modulation frequency on the axis are normalized by the natural linewidth $\Gamma=6 \rm{MHz}$, and the flux is normalized with the maximum flux when scanning the parameters. }
\end{figure}

The widened frequency spectrum has the potential to increase the capture velocity of cooling lasers, thereby enhancing the beam flux \cite{Li2015, Barbiero2020, Lee:17}. In our experiments, the flux was systematically scanned at different carrier and modulation frequencies. Fig.~\ref{fig7} presents a contour map illustrating the beam flux as a function of the carrier frequency (detuning) and the modulation frequency of the phase-modulated cooling laser. A RF signal provides 0.25 modulation depth of the cooling laser. The whole optical intensity of the cooling laser is $23.38\,\rm{mW/cm^2}$ ($14I_0$), and the optical intensity of the pushing laser is $2.55\,\rm{mW/cm^2}$ ($1.52I_0$). The pushing laser is not modulated in this experiment. The results reveal a maximum flux at a modulation frequency near 0 Hz, indicating that the white-color cooling laser has no enhancing effect on  the flux of the 2D+MOT beam.This observation may be attributed to collisions between the atoms and the vacuum chamber, leading to alterations in the velocities of the atoms in each cooling direction. Consequently, atoms with a large Doppler detuning may resonate with the cooling laser after such collisions. Another consideration is related to atoms ejected from the trap: multiple-sidebands cooling might increase losses due to fine-structure changing collisions. This could pose challenges in trapping large-mass alkali-metal atoms like potassium, rubidium, and cesium when employing multiple-sidebands cooling, as suggested in previous studies \cite{Li2015}.

\section{Conclusion}

We present a cold atomic beam source based on a 2D+MOT, generating
a continuous cold beam of ${}^{87}$Rb atoms with a flux of reaching $4.3\times10^9 \, \rm{s^{-1}}$, a mean velocity of $10.96(2.20)\,\rm{m/s}$, and a transverse temperature of $16.90(1.56)\,\rm{\mu K}$. The mean velocity of the cold atoms can be adjusted from $9.5\,\rm{m/s}$ to $14.6\,\rm{m/s}$ by phase-modulating the pushing laser. This cold atomic beam source is suitable for application in atom interferometers and clocks, offering continuous operation, a slow and tunable atomic beam velocity, and a narrow transverse distribution that enhances  the bandwidth, sensitivity, and signal contrast of these devices.

For an atom interferometer with an ideal fringe contrast of C=1, the quantum projection noise-limited phase noise is $\Delta \Phi \approx 15.25\,\rm{\mu rad/ \sqrt{Hz}}$ \cite{Itano1993}. This result translates to a short-term rotation rate sensitivity of $\Delta \Omega \approx 5.24 \times 10^{-8}\,\rm{(rad/s)/\sqrt{Hz}}$ or a short-term acceleration noise of $\Delta a \approx 1.05 \times 10^{-6}\,\rm{(m/s^2)/\sqrt{Hz}}$ for an atom interferometer-based gyroscope or accelerometer employing this source.

We studied the flux dependence on cooling laser intensity at different repumping laser intensities and quantitatively determined the repumping laser requirements for the 2D+MOT. The experimental results are consistent with simulations derived from the cooling efficiency-modified model. Future efforts should concentrate on further suppressing the longitudinal velocity distribution using techniques such as  optical molasses while maintaining low transverse temperatures. Furthermore, there is potential for enhancing both flux and reducing temperatures by optimizing the magnetic field and cooling beam profiles.

\begin{backmatter}
\bmsection{Funding}
National Natural Science Foundation of China (Grant No.61473166).
\bmsection{Acknowledgments}
This work was supported by the National Natural Science Foundation of China (Grant No.61473166). 
\bmsection{Disclosures}
The authors declare no conflicts of interest.
\bmsection{Data Availability Statement}
Data underlying the results presented in this paper are not publicly available at this time but may be obtained from the authors upon reasonable request.
\end{backmatter}

\bibliography{sample}

\end{document}